\documentclass[grl]{agu2001n}
\usepackage{graphicx}
\authorrunninghead{L\'OPEZ-RUIZ ET AL.}

\titlerunninghead{A MODEL OF CHARACTERISTIC EARTHQUAKES}

\authoraddr{J. B. G\'omez,
Departmento de Ciencias de la Tierra, Universidad de Zaragoza,
Pedro Cerbuna 12, Facultad de Ciencias, Edificio C, 50009
Zaragoza, Espa\~na. (jgomez@unizar.es)}
\begin{document}

%
%

\title{Asymptotic Behavior of a Model of Characteristic Earthquakes and its
Implications for Regional Seismicity}
%

%
%


\author{Ricardo L\'opez-Ruiz}
\affil{Department of Computer Sciences and BIFI, University of
Zaragoza, Spain}

\author{Miguel V\'azquez-Prada}
\affil{Department of Theoretical Physics, University of Zaragoza,
Spain}

\author{Javier B. G\'omez}
\affil{Department of Earth Sciences, University of Zaragoza,
Spain}

\author{Amalio F. Pacheco}
\affil{Department of Theoretical Physics and BIFI, University of
Zaragoza, Spain}

\begin{abstract}
The recently introduced Minimalist Model \markcite{[{\it
V\'azquez-Prada et al,} 2002]} of characteristic earthquakes
provides a simple representation of the seismicity originated in a
single fault. Here, we first characterize the properties of this
model for large systems. Then, assuming, as it has been observed,
that the size of the faults in a big enough region is  fractally
distributed, with a fractal dimension $D=1.7$, we describe the
total seismicity produced in that region. The resulting catalogue
accounts for the addition of all the characteristic events
triggered in the faults  and also includes the rest of
non-characteristic earthquakes. The global accumulated
size-frequency relation correctly describes a Gutenberg-Richter
form, with a $b$ exponent of around 0.7.
\end{abstract}

%
%

%

\begin{article}

%
%


%
%

\section{Introduction}
In regional and global seismicity, there is a well established
fractal relation, the Gutenberg-Richter law, that can be expressed
in the form

 \begin{equation}
 \dot{N} \propto S^{-b}, \label{GR}
 \end{equation}
where $\dot{N}$ is the number of observed earthquakes in a year
with rupture area greater than $S$, and $b$ is the so-called
$b$-value, which is in the range 0.5-1.5. The Gutenberg-Richter
law implies that earthquakes are, on a regional or global scale, a
self-similar phenomenon without any characteristic length scale.

It is important to notice, however, that the Gutenberg-Richter law
is a property of regional seismicity, appearing when we average
seismicity over big enough areas and long enough time intervals.
In the last ten years, a wealth of data has been collected to
extract statistics on individual systems of earthquake faults
\markcite[{\it Wesnousky}, 1994; {\it Sieh}, 1996; {\it Petersen
et al.}, 1996]. Interestingly, it has been found that the
distribution of earthquake magnitudes may vary substantially from
one fault to another and that, in general, this type of
size-frequency relationship is different from the
Gutenberg-Richter law. Many single faults or fault zones display
power-law distributions only for small events (small compared with
the maximum earthquake size a fault can support, given its area),
which occur in the intervals between roughly quasi-periodic
earthquakes of much larger size which rupture the entire fault.
These large quasi-periodic earthquakes are termed
``characteristic'' \markcite[{\it Schwartz and Coppersmith} 1984],
and the resulting size-frequency relationship, Characteristic
Earthquake distribution.

With the purpose of representing the seismicity originated in a
single fault, we have recently introduced a simple model called
``Minimalist'' \markcite[{\it V\'azquez-Prada et al, 2002}]. In
this model, a one-dimensional vertical array of length $N$ is
considered. The ordered levels of the array are labelled by an
integer index $i$ that runs upwards from 1 to $N$. This system
performs two basic functions: it is loaded by receiving stress
particles in its various levels and unloaded by emitting groups of
particles through the  first level $i=1$. These emissions which
relax the system are considered to be earthquakes, and the number
of consecutively empty lower levels, $k$, in each event is assumed
to be equal to the ruptured area.

The main goal of this paper is the study of the  total seismicity
produced in a region, using the Minimalist Model for the
description of the seismicity coming from each individual fault.
Thus in the coming paragraphs, we will briefly review some
properties of the Minimalist Model, and will identify its
asymptotic ($N \rightarrow \infty$) behavior. This information,
crossed with the observed fractal distribution  of fault sizes at
a regional level, will give us a size-frequency distribution for
the earthquake  population which will be discussed and compared
with Eq. (1).

\section{The Minimalist Model and its Asympt\-otic Behavior}

In the Minimalist Model, the rules  for the loading and relaxing
functions in the system are:

\begin{enumerate}

\item The incoming particles arrive at the system at a constant
time rate. Thus, the time interval between each two successive
particles will be considered the basic time unit in the evolution
of the system.

\item All the positions in the array, from $i=1$ to $i=N$, have
the same probability of receiving a new particle. When a position
receives a particle we say that it is occupied.

\item If a new particle comes to a level which is already
occupied, this particle has no effect on the system. Thus, a given
position $i$ can only be either non-occupied when no particle has
reached it, or occupied when one or more particles have.

\item The level $i=1$ is special. When a particle goes to this
first position a relaxation event occurs. Then, if all the
successive levels from $i=1$ up to $i=k$ are occupied, and
position $k+1$ is empty, the effect of the relaxation --or
earthquake-- is to unload all the levels from $i=1$ up to $i=k$.
Hence, the size of this relaxation is $k$, and the remaining
levels $i>k$ remain unaltered in their occupancy.

\end{enumerate}

 Thus, we notice that this model was devised in a spirit akin
to the Sand-Pile model of Self- Organized Criticality: physics is
not apparent, but the model is able to grasp the basics of the
routine of a fault dynamics. In this case, the presence of an
asperity controlling the mechanism of relaxation in the system, is
a necessary ingredient of this cellular automaton.

From what has been mentioned above, this model has no parameter;
the size $N$ is the unique specification to be made, and the
spatial correlation is induced by rule 4 above. Now, the state of
the system is given by stating which of the ($i>1$) $N-1$ levels
are occupied. Each of these states corresponds to a stable
configuration, and therefore the total number of possible
configurations is $2^{(N-1)}$. We use the term ``total occupancy''
for the configuration in which all levels except the first are
occupied.

As this model is one-dimensional, extensive Monte Carlo
simulations can be performed to accurately explore its properties.
It can also be studied, for small system sizes, from the
perspective of Markov Chains \markcite[{\it V\'azquez-Prada et
al}, 2002].

The results for the earthquake size-frequency relation, $p_k$, are
drawn in Figure 1. In this figure, there are two notable
properties to be commented on. First and most important, we see
that the characteristic relaxation, $k=N$, has a much higher
probability of occurrence than the big relaxations but with $k<N$.
In fact, for $N=10$, $100$ and $1000$, the probability of these
three characteristic relaxations differs very little, and is about
$10\%$. We can express this fact by saying that in this model,
{\it grosso modo}, one would likely observe only very small
earthquakes and the characteristic one. And secondly, we observe
the perfect coincidence of these curves of probability for systems
of different size $N$.

This second property of the model is related to what we will call
henceforth  the ``tail splitting'' property , which is expressed
by the relation
\begin{equation}
P_N (N) = P_{N+1} (N+1) + p_N (N+1). \label{PN}
\end{equation}
The meaning of this expression is as follows. The probability of
occurrence of a characteristic earthquake in a system of size $N$
is equal to the sum of the probability of a characteristic in a
system of size $N+1$ plus the probability of a non-characteristic
of size $N$ in a system of size $N+1$. Thus, capital letter $P$ is
reserved for the probability of a characteristic event, and small
$p$ for the probability of non-characteristic earthquakes, which
are independent of the system size.

This ``tail splitting'' property holds in a family of models that
include the Minimalist Model. Its consequence is that if one knows
$p_k(N)$ for all $1 \le k < N$, then one knows $P_M(M)$ for $1
\le M \le N$. And vice versa, using Eq. (\ref{PN}), the knowledge
of the $P$s leads to the knowledge of the $p$s. In the Minimalist
Model, the value of these two sets of probabilities can be easily
obtained, for small $N$, by means of diagonalizations of the
corresponding Markov matrices, and in general by Montecarlo
simulations. For large systems one expects a regular behavior in
the decaying rate of these probabilities. This information is
shown in Fig. 2. In Fig. 2a, the values of the $P$s are fit by the
function
\begin{equation}
P_N (N) = c(\log N)^{-\alpha},\label{asymp}
\end{equation}
where $\alpha=0.95$ and $c$ is a constant. Eq. (\ref{asymp}) means
that the probability of occurrence of a characteristic earthquake
goes to zero as the system size goes to infinity, although very
slowly. For example, for  a system of size $N=1000$ the
probability of a characteristic earthquake is $P_{1000}(1000) =
0.07582$; for a system size $N=10^6$, $P_{10^6}(10^6) = 0.04662$;
and for a system size $N=10^8$, $P_{10^8}(10^8) = 0.02986$. A
difference of five orders of magnitude in system size produces a
drop in the probability of a characteristic earthquake from 7.5\%
to 3\%. Equating the rupture area of an earthquake with the
parameter $N$ in the minimalist model, a difference of five orders
of magnitude in rupture area is roughly that which exists between
earthquakes of magnitude 3 and 8.

As a verification of the goodness of this fit, in Fig. 2b we have
plotted the ratio $P_N(N)/P_{N'}(N')$ of the probability of a
characteristic earthquake for two different system sizes ($N$ and
$N'$) against the ratio $(\log N'/\log N)^\alpha$ for
$\alpha=0.95$. If Eq. (\ref{asymp}) does indeed describe well the
asymptotic behavior of the probability of occurrence of
characteristic earthquakes in the minimalist model, then this plot
should be a straight line of slope 1. The points are colour-coded
with respect to the offset between the values of $N$ and $N'$ used
to calculate the ratios $P_N(N)/P_{N'}(N')$ and $(\log N'/\log
N)^\alpha$. An offset of one means that the ratio is calculated
for two consecutive data points. An offset of two is for ratios
calculated for next-nearest neighbor data points, and so on. As
can be seen, all the data points, regardless of the offset, fall
on the slope-one straight line, indicating that function (1) is a
good description of the asymptotic behavior. To further check that
the value $\alpha=0.95$ is statistically different from
$\alpha=1$, in Fig. 2c we have plotted again the same ratios as in
Fig. 2b but with an exponent $\alpha=1$. The clear departure from
the slope-one straight line for big offsets shows that the correct
exponent is $\alpha=0.95$ and not $\alpha=1$.

Now, using Eqs. (\ref{PN}) and (\ref{asymp}), always for large
$N$, we have
\begin{equation}
c(\log N)^{-\alpha} = c(\log(N+1))^{-\alpha} + p_N(N+1)
\end{equation}
and therefore
\begin{equation}
 p_N(N+1) = \frac{c \alpha}{N(\log N)^{\alpha+1}}.
\end{equation}

\section{Implication for the global seismicity and discussion}

Observational data seem to show that the frequency-size
distribution of faults is fractal \markcite[{\it Hirata}, 1989,
{\it Barton et al.}, 1996]. Specifically, the results of
\markcite{{\it Barton et al.} (1995)} in Nevada indicate that the
fractal dimension has a mean value of $D = 1.7$. Using this
information, our result for the probability of occurrence of an
earthquake of size $k$ originated in any of the faults of the
region is
\begin{equation}
p_k \approx \sum^\infty_{N=k+1} N^{-D} \frac{c\alpha}{k (\log
k)^{\alpha+1}} + k^{-D} \frac{c}{(\log k)^\alpha}. \label{pk1}
\end{equation}
In this formula, the first term  accounts for  the contribution of
the non-characteristic earthquakes originated in the faults, and
the second corresponds to the contribution of only the
characteristic ones. Eq. (\ref{pk1}) is equivalent to
\begin{equation}
p_k \approx \frac{c\alpha}{k (\log k)^{\alpha+1}}
\sum^\infty_{N=k+1} N^{-D} + k^{-D} \frac{c}{(\log k)^\alpha}.
\label{pk2}
\end{equation}
After computing the sum in the first term, we obtain
\begin{equation}
p_k \approx \frac{c\alpha}{k (\log k)^{\alpha+1}}
\frac{k^{1-D}}{D-1} +k^{-D} \frac{c}{(\log k)^\alpha}. \label{pk3}
\end{equation}
Rearranging this equation, we get

\begin{equation}
p_k \approx k^{-D} \frac{c}{(\log k)^\alpha} \left( 1 +
\frac{\alpha}{D-1} \frac{1}{\log k} \right). \label{pk4}
\end{equation}
When $k$ is large, Eq. (\ref{pk4}) simplifies to
\begin{equation}
p_k \propto k^{-D} \frac{1}{(\log k)^\alpha}. \label{pk5}
\end{equation}

In other words, the net result is equivalent to the product of a
first factor coming from the fault size distribution times a
second factor coming from the seismicity of the individual faults.
Note that the impact of the second factor, i.e.  the logarithmic
term, in the decay rate of $p_k$,  is very light in comparison to
the first power-law factor coming from the assumed fault
distribution. This is put in evidence in Fig. 3a, where
simulating the regional seismicity, we have drawn all the
earthquakes resulting from a large number of minimalist systems,
acting simultaneously, and distributed  in size  in the fractal
form $N^{-D}$ with $D=1.7$. In Fig.3b, we have reorganized the
information of Fig.3a, in an accumulated distribution for a more
direct comparison with Eq. (\ref{GR}). As shown in this figure,
our result corresponds to a Gutenberg-Richter relation with an
exponent around 0.75.

Hence, our main conclusion in this paper is that, using this
model, the slope $b$ appearing in the Gutenberg-Richter graph of
the regional seismicity is basically due to the effect of the
fractal distribution of fault sizes.

\setcounter{equation}{0}

%
%


%
%


%
%

\begin{acknowledgments}
This work was supported in part by the Spanish DGICYT (Project
BFM2002-01798).
\end{acknowledgments}

%
%

\clearpage

\begin{figure}
\noindent\includegraphics[width=20pc]{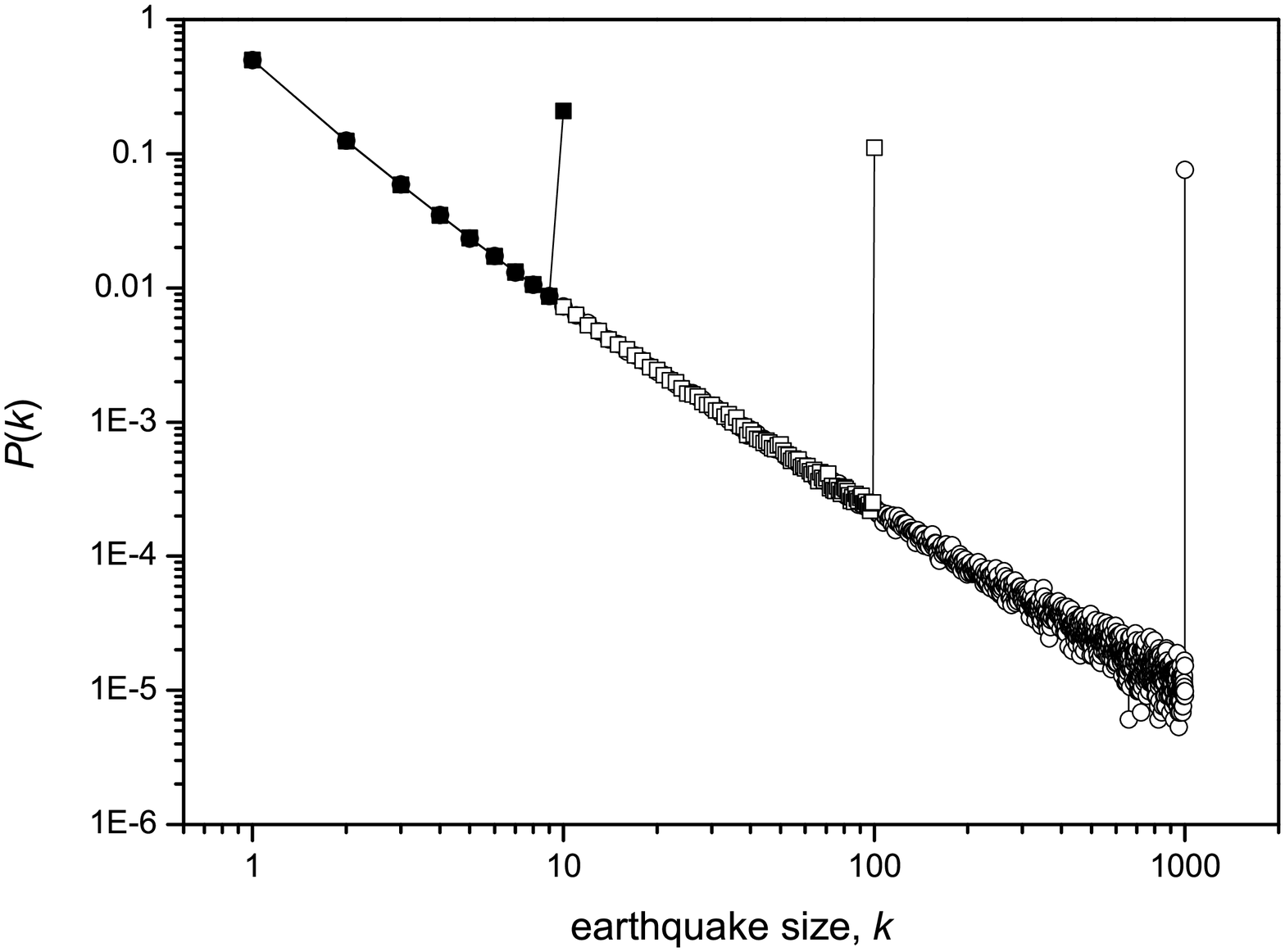}
 \caption{Probability of occurrence of earthquakes of size $k$
 in the Minimalist Model. Three simulations are superimposed,
 corresponding to systems with $N=10$, $N=100$, and $N=1000$ elements. }
\end{figure}

\clearpage

\begin{figure}
\noindent\includegraphics[width=16pc]{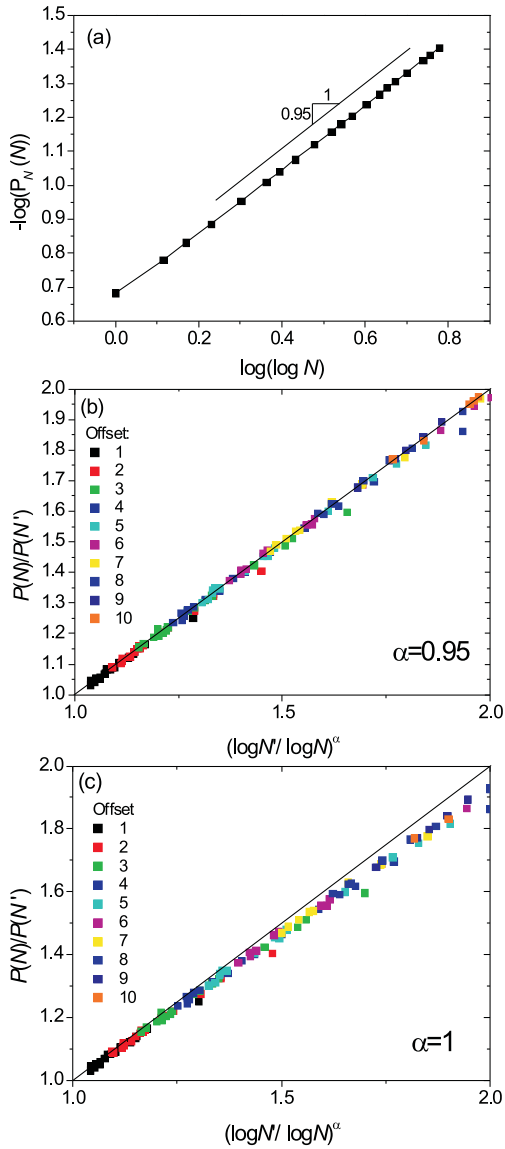}
 \caption{Asymptotic behavior of the Minimalist Model. (a)
 Logarithmic-decaying fit to the probability of occurrence of the
 characteristic earthquake in a system of size $N$. (b) Plot of the ratio
 $P_N(N)/P_{N'}(N')$ against the ratio $(\log N'/\log
N)^\alpha$, for $\alpha=0.95$, where $N$ and $N'$ are two
different system sizes. If the logarithmic-decaying fit is
correct, all the data points in this graph should plot on the
diagonal straight line of slope one, as is the case. (c) Same as
(b), but with $\alpha=1$ to show that the value $\alpha=0.95$ is
statistically different from $\alpha=1$. In this case the
large-offset data points plot off the slope one line. }
\end{figure}

\clearpage

\begin{figure}
\noindent\includegraphics[width=16pc]{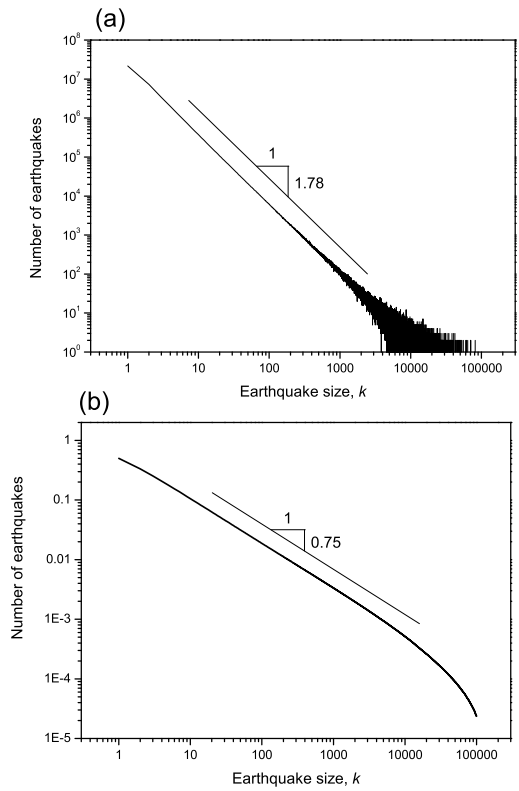}
 \caption{(a) Discrete Gutenberg-Richter relation for earthquakes occurring in
 minimalist-model faults with a power-law (fractal) size-frequency
 distribution with fractal dimension $D=1.7$. The slope of
 the straight line is $-1.78$. (b) Accumulated Gutenberg-Richter relation for earthquakes 
occurring in
 minimalist model faults with a power-law (fractal) size-frequency
 distribution. The slope of the straight line is $-0.75$.}
\end{figure}

%
%
%

%
%
%
%
%
\end{article}

\end{document}